\documentclass[%
 reprint,
 showpacs,preprintnumbers,
 amsmath,amssymb,
 aps,
prb
]{revtex4-1}
\usepackage{dcolumn}
\usepackage[dvipdfmx]{graphicx}
\usepackage{subfigure}
\usepackage{color}
\usepackage{mathrsfs}

\makeatletter
\def\btt#1{\texttt{\@backslashchar#1}}
\DeclareRobustCommand\bblash{\btt{\@backslashchar}} \makeatother

\renewcommand{\Vec}[1]{\mbox{\boldmath$#1$}}

\def\Vec#1{\textbf{#1}}
\def\gsim{\lower -0.3ex \hbox{$>$} \kern -0.75em \lower 0.7ex
	\hbox{$\sim$}}
\def\lsim{\lower -0.3ex \hbox{$<$} \kern -0.75em \lower 0.7ex
	\hbox{$\sim$}}

\begin{document}

\title{Dirac electrons on three-dimensional 
graphitic zeolites\\
--- a scalable mass gap}

\author{Mikito Koshino}
\affiliation{Department of Physics, Tohoku University, Sendai 980-8578, Japan}

\author{Hideo Aoki}
\affiliation{Department of Physics, University of Tokyo, Hongo, Tokyo 113-0033, Japan}

\date{\today}

\begin{abstract}
A class of graphene wound into three-dimensional periodic 
curved surfaces 
(``graphitic zeolites") is proposed and their 
electronic structures are obtained to 
explore how the massless Dirac fermions behave on 
periodic surfaces. 
We find in the tight-binding model 
that the low-energy band 
structure around the charge neutrality point 
is dominated by the topology (cubic or gyroid) 
of the periodic surface as well as by the spatial period $L$ 
in modulo 3 in units of the lattice constant. 
In both of cubic and gyroid cases the Dirac electrons become massive around the 
charge neutrality point, where the band gap is shown to 
scale as $1/L$ within each mod-3 class.  
Wave functions around the gap 
are found to have amplitudes sharply peaked around 
the topological defects that are required to 
deform the graphene sheet into a three-dimensional 
periodic surface, and this is shown to 
originate from non-trivial Bloch phases at $K$ and $K'$ points of the original graphene.
\end{abstract}


\maketitle

\begin{figure}
 \begin{center}
 	\includegraphics[width=0.9\hsize]{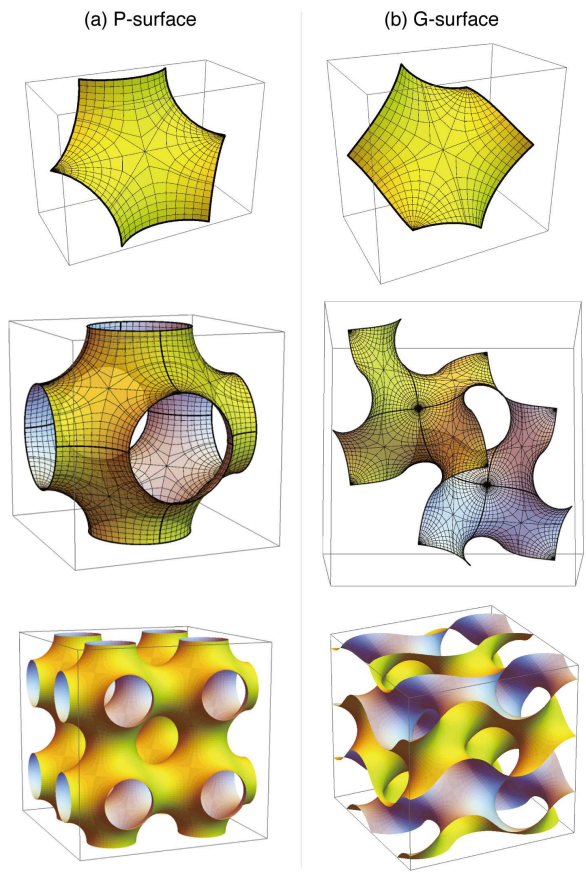}
\end{center}
 	\caption{
Structures of continuous (a) P-surface and (b) G-surface. 
In each column, top panel: a primitive surface patch, 
middle: 3D unit structure, 
bottom: periodic connection of the units.  
}
\label{fig_p_and_g}
 \end{figure}

{\it Introduction---} 
While a vast variety of phenomena 
arising from massless 
Dirac electrons
have been revealed for two-dimensional (2D) graphene, 
a natural and important question is 
what will happen to the electronic structure 
if three-dimensional (3D) structures are constructed from graphene.
Specifically, when we envisage 
structures that are periodic in each of $x, y$ and $z$ 
directions, we are talking about triply-periodic 
surface, or ``graphitic zeolites", where we have 
a 3D network of graphene sheets.  
Interests are, first, band-theoretically how the massless Dirac particles 
behave in totally different spatial topologies\cite{helicoid1}. 
Second, when the flat honeycomb lattice of graphene 
is wound into such 3D geometries, we have to 
introduce topological defects, 
and the electronic properties of carbon systems 
in such cases have been studied in the context of fullerenes, carbon nanotube caps and graphene cones.\cite{Kroto85,Iijima91,Gonzalez92,Gonzalez93,Tamura94,Tamura97,Lammert04}  
For a 3D graphitic zeolite, we 
have to introduce topological defects as more than 
six-membered rings, around which the 
surface has negative Gaussian curvatures,\cite{Terrones91,Terrones92,Vanderbilt92,Lenosky92}  
as illustrated in Fig.\ \ref{fig_p_and_g}. 

Thus the question we pose here is: In such 3D graphene 
structures, 
how its low-energy spectrum would look like --- can it still be a 3D massless Dirac electron, or otherwise, 
especially around the ``charge neutrality point" 
(originally the Dirac point of flat graphene)?   
The question is not only academic, but should be relevant 
to the electronic properties
in 3D carbon nanostructures that are realized in recent experiments.\cite{Ma2001,Ma2002,kyotani2009,Li2013,Wu2013,Xu2013,WChen2011,Yang2013,ZChen2011,Cao2011,Ito2014}  
Specifically, a high-quality 3D graphene network
was very recently fabricated with a nanoporous template material,\cite{Ito2014}
where the system was found to comprise interconnected single graphene sheet 
with structural length scale of 100 nm to 1 $\mu$m.  While the 
samples fabricated are not periodic, they are expected to retain 
the low-energy characteristics of graphene owing to a high quality, so that they should serve as a 
starting point for realizing massless Dirac particles on curved surfaces.  

Theoretically, band structures of periodic curved surfaces have been 
studied for conventional (nonrelativistic) electrons 
for various surfaces including Schwarz's P (simple 
cubic) and D (diamond-structured) surfaces\cite{Aoki2002,Aoki2004} and also G (gyroid) surface.\cite{KoshinoAoki2005}  
The massless Dirac particle on 
periodic curved surfaces poses an entirely different problem.  
A crucial interest is the fate of the Dirac cone in the band structure in these structures. 
Possible structures of 3D periodic graphene were first theoretically proposed \cite{Terrones91,Terrones92,Vanderbilt92,Lenosky92,Fujita95, Weng2014, Tagami2014}, 
and band structures were obtained for some 
specific cases with relatively small unit-cell sizes of the order of the atomic scale.
However, to consider systematics of the Dirac particle on  periodic curved surfaces 
on a more general ground, particularly around the 
charge neutrality point, 
we need to look at 
the energy band structure not only for 3D structures with different topologies, but 
for different length scales in the periodic structures 
to fathom the scaling behavior of the band structure and mass gap.  

\begin{figure}
 \begin{center}
 	\includegraphics[width=0.9\hsize]{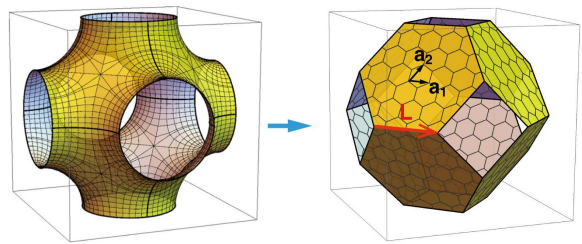}
\end{center}
 	\caption{
Construction of zeolitic graphite in P-surface 
geometry (left panel) as a network of honeycomb sheet 
 (right), where the primitive patch is taken to be a 
 graphene fragment with a hexagonal shape characterized 
 by ${\bf L}$.  ${\bf a}_1, {\bf a}_2$ are primitive 
 vectors for honeycomb. 	}
\label{fig_unit_patch}
 \end{figure}

These have motivated us to study 
in this paper the electronic energy spectra of 3D graphene networks 
with varied spatial periods and structures.  
As the surface topology, we consider two typical periodic minimal surfaces, 
which are called Schwarz's P-surface (cubic) 
and G-surface (gyroid with helical symmetry along 
each of $x,y,z$ directions) as shown in Fig.\ \ref{fig_p_and_g}(a) and (b), respectively.  
We construct the corresponding honeycomb networks having these topologies 
as illustrated in Fig.\ \ref{fig_unit_patch}, and calculate the band structure with the simple tight-binding model.  
By varying the periodicity of these 3D structures 
we study the scaling behavior of the band structure around the charge neutrality point.  

We shall first find that the energy spectrum around the charge neutrality point is 
essentially determined by the surface topology (P or G).
In each case we reveal a ``mod-3 rule", where the spatial period 
in modulo 3 in units of the original honeycomb lattice constant dominates the bands.  
The systems in the same category (with the same topology and the same mod-3 number) 
have almost identical low-energy band structures, where 
energy (band width, etc) scales as $1/L$, where $L$ is the spatial period.
In each case, the original Dirac cone becomes massive with the mass gap also scaling as $1/L$.
Thus we conclude that massless Dirac electron bound on the 3D periodic surfaces 
behaves as a {\it massive Dirac electrons with a scalable mass}.
We also find the wave functions around the gap 
have singularily large amplitudes concentrated around 
the topological defects that are required to make 
the system periodic in 3D.
 
{\it Formulation---} 
P-surface (space group: Im3m) and G-surface (Ia$\bar{3}$d) as networks of continuous membrane are depicted in 
Fig.\ref{fig_p_and_g} (a) and (b), respectively.  
We can construct the 3D periodic structure 
from a primitive patch (top panel) 
to construct the 3D unit structure composed of eight primitive patches (middle panel), which are then 
periodically connected (bottom).  
Zeolitic graphites having the topologies of P or G-surfaces can then be constructed by taking a  
graphene fragment with a hexagonal shape 
as the primitive patch, 
as illustrated in Fig.\ \ref{fig_unit_patch} for P-surface.
We can characterize the graphene fragment with 
indeces $(m,n)$, which specify 
the lattice vector connecting the corners of the hexagon,
\begin{align}
{\bf L} = m {\bf a}_1 + n {\bf a}_2,
\label{eq_L}
\end{align}
where ${\bf a}_1=a(1,0)$ and ${\bf a}_2=a(1/2,\sqrt{3}/2)$ are primitive unit vectors
of the original graphene honeycomb lattice.  
The resulting 3D structure has an eight-membered ring
at each vertex where four patches meet,
while it is composed of six-membered rings elsewhere. 
We can model the system 
with the simple tight-binding Hamiltonian for $\pi$ orbital of carbon,  
$H= -t \sum_{i,j} c^\dagger_i c_j$, 
where $t\simeq 3$ eV is the hopping integral, $c^\dagger_i$ creates an electron on $i$-th site, 
and $i,j$ are summed over nearest-neighbor sites. 
We neglect the curvature effect on the hopping parameter $t$, and 
take it as a constant and the unit of energy ($t\simeq 3$ eV).
Effects of the curvature of the structure will be discussed below.  

\begin{figure}
 \begin{center}
 	\includegraphics[width=1.\hsize]{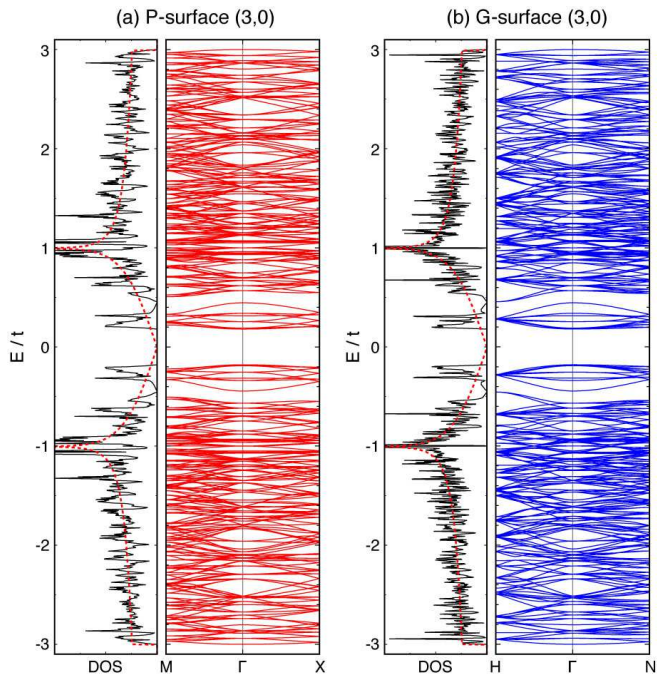}
\end{center}
\caption{Band structure (right) and the density of states (left)
for (a) P-surface graphene and (b) G-surface graphene, 
both illustrated for $(m,n)=(3,0)$.
Dashed curves in the density of states indicate the 2D graphene's. 
}
\label{fig_band_p30_and_g30}
 \end{figure}

 \begin{figure}
 \begin{center}
 	\includegraphics[width=1.\hsize]{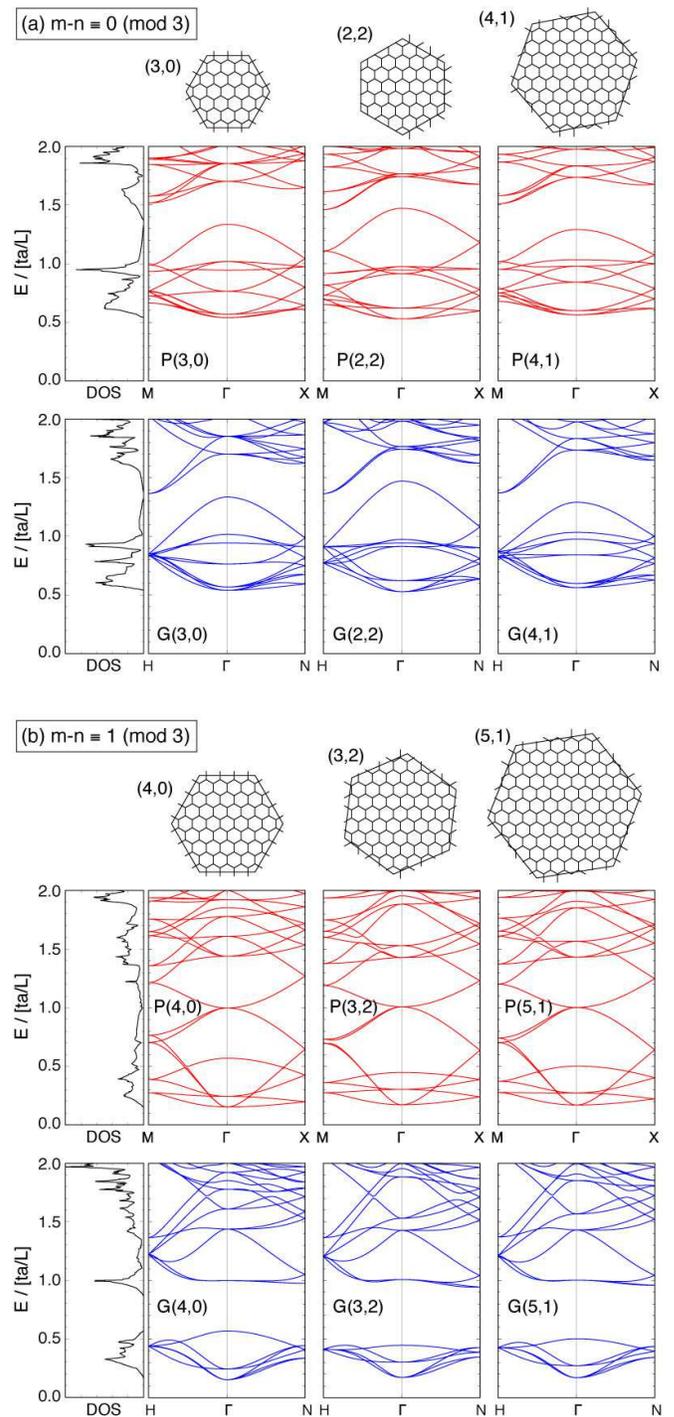}
\end{center}
 \caption{
 Band structures of P-surface (middle panel) and G-surface (bottom) graphenes 
(blow up of the positive-energy regions just above the charge neutral point) 
 for (a) $m-n\equiv 0$ (mod 3) or (b) $m-n\equiv 1$.  
Three columns represent the results for 
various values of $(m,n)$, with 
the top inset showing the primitive graphene fragment. 
The energy is rescaled 
into a dimensionless $E/(ta/L)$, and 
the density of states is shown for the smallest period 
in each of (a), (b).
 		 	}
\label{fig_gp_0_and_1}
 \end{figure}
  
\begin{figure}
 \begin{center}
 	\includegraphics[width=1.\hsize]{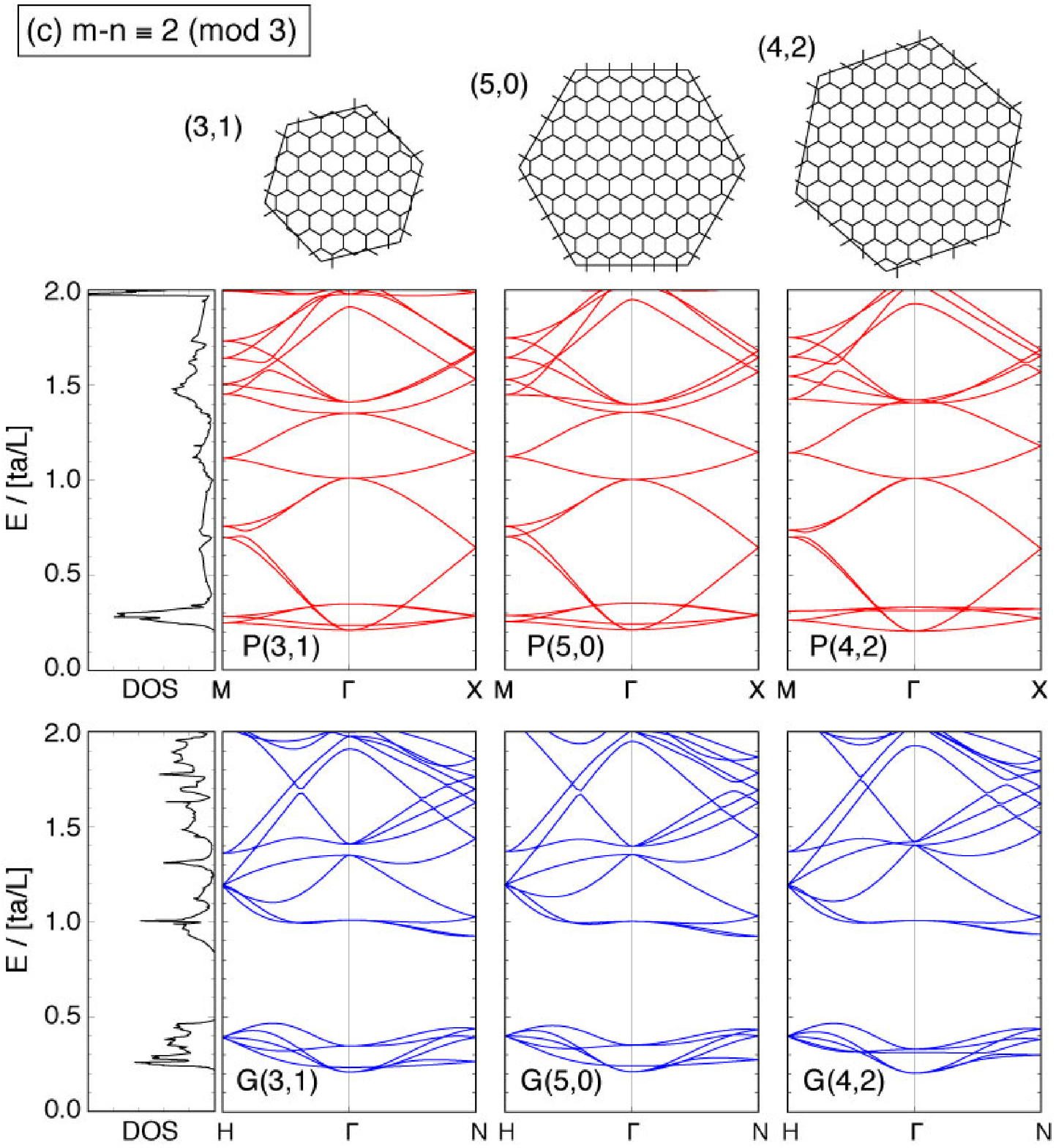}
\end{center}
 \caption{A plot similar to Fig.\ \ref{fig_gp_0_and_1} for $m-n\equiv 2$ (mod 3).
 		 	}
\label{fig_gp_2}
 \end{figure}

{\it Results---} 
Fig.\ \ref{fig_band_p30_and_g30} shows a typical band structure for (a) P-surface graphene
and  (b) G-surface graphene, here illustrated for 
$(m,n)=(3,0)$. 
The spectrum is electron-hole symmetric with respect to $E=0$ (the charge neutrality point) 
in the present tight-binding model. 
Due to the band folding coming from the 3D periodicity, 
we have many bands, and if we look at the 
region around the charge neutrality point, 
the density of states (DOS) drastically 
deviates from graphene's $E-$linear behavior (dashed curve).  
Specifically, the energy band is gapped around $E=0$ 
at $\Gamma$, 
and some clustering is seen in the conduction (valence) bands just 
above (below) the gap.

A fuller picture is shown for $m-n\equiv 0$ (mod 3) in 
Fig.\ \ref{fig_gp_0_and_1}(a),  $m-n\equiv 1$ 
(Fig.\ \ref{fig_gp_0_and_1}(b)) and  
$m-n\equiv 2$ (Fig.\ref{fig_gp_2}), for P-surface graphene (middle panels) and for G-surface graphene (bottom). 
For each case of $m-n$ (mod 3), we plot the results for three 
different choices of $(m,n)$ with the top inset showing the the primitive graphene fragment. 
Throughout, the energy is taken to be a dimensionless one, $E/(ta/L)$.
This choice comes from our expectation that 
the low-energy spectrum around $E=0$ can be well described
by the continuum massless Dirac field when the length scale of the 3D network is much longer than the atomic scale. 
Then the energy spectrum should be scaled as $1/L$. 

We clearly see that the rescaled band structures belonging to the same family 
[i.e., the same  $m-n$ (mod 3) and the same surface topology (P or G)] 
are similar to each other, while distinct between different families.  
Specifically,  the Dirac electron becomes massive, where the mass 
scales as $\sim1/L$. 
The modulo-3 rule is reminiscent of the situation in carbon nanotubes\cite{Ando2005},
and it is in fact an intrinsic property of graphitic systems having closed loop structures.
In the flat graphene, the low-energy wave functions have a character of 
the states around the Brillouin zone corners $\Vec{K}_\pm
\equiv 2\pi/a(\pm 2/3,0)$,
and when graphene is rolled into a carbon nanotube with circumference vector $\Vec{t}$,
the wave function acquires a 
phase $e^{i\Vec{K}_\pm \cdot\Vec{t}}$. 
Similarly, the 3D graphene network has a 
periodic array of close loops 
composed of hexagon's sides,
where similar phase factors emerge.
The phase factor for a single side of hexagon [with the length $\Vec{L}$ defined by Eq.\ (\ref{eq_L})]
amounts to $e^{i\Vec{K}_\pm\cdot \Vec{L}} = e^{\mp 2\pi i (m-n)/3}$, which is indeed determined 
solely by $m-n$ (mod 3),
and explains the modulo-3 rule in the band structure.
This situation around the Dirac point 
sharply contrasts with the those around 
the bottom (or top) of the whole energy spectrum (i.e. $E/t \approx \pm 3$), 
where we can show that the rescaled band structure only depends on the 
surface configuration (P or G), and does not 
depend on $(m,n)$.  
This is because the bottom (or top) region corresponds to 
$\Gamma$ point of the original graphene,\cite{Aoki2002} where phase factors do not appear in the continuum 
model, and the characteristic energy scale is given by $\hbar^2/(m^* L^2)$
with $m^*$ being the $\Gamma$-point effective mass. 

If we now turn to the G-surface graphene, 
there is a specific interest as follows.  
In this ``gyroid" structure, the surface has a screw axis 
along each of $x, y$ and $z$ directions.  In fact 
the G-surface is the only triply-periodic surface 
having this property.  The band structure of 
electrons has been analyzed by the present authors with 
Schr\"{o}dinger equation \cite{KoshinoAoki2005}.  
There, we found that an effect of the helical structure appears as the multiple band sticking on the Brillouin zone boundaries.  
If we look at the present result for Dirac particles on G-surface, 
we immediately notice a salient feature that we have also 
{\it multiple band sticking}, 
which occurs for the Dirac particles at 
H point in the Brillouin zone and for the bunch of 
energy band just above the mass gap.  
In terms of the space group, this feature is expected to arise from a general property of non-symmorphic cases 
that promotes band touching \cite{Bradley1972,Parameswaran2013}.

\begin{figure}[ht]
 \begin{center}
 	\includegraphics[width=1.\hsize]{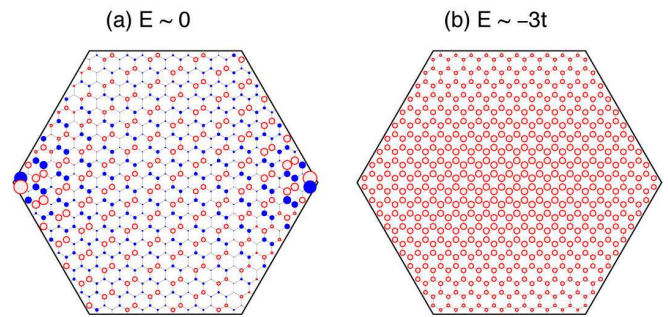}
\end{center}
 \caption{
Typical wave functions 
around (a) the charge neutrality point $(E\simeq 0)$, 
and (b) the bottom $(E\simeq -3t)$ of the whole energy spectrum, 
both at $\Gamma$-point in the 3D Brillouin zone.  
Here $(m,n)=(10,0)$, and the radius of each circle indicates the amplitude while open and solid circles represent sign of the wave function.
 		 	}
\label{fig_wave}
 \end{figure}

The wave functions also exhibit a notable feature.   Figure \ref{fig_wave} 
plots typical wave functions, here for $(m,n)=(10,0)$, 
near (a) the charge neutrality point $(E\simeq 0)$ 
and near (b) the band bottom $(E\simeq -3t)$, both at $\Gamma$-point in the 3D Brillouin zone.  
We can see that the wave function for $E\simeq 0$ has amplitudes sharply 
concentrated around the topological defects 
(which correspond to the corners of the patches), 
whereas that for  $E\simeq -3t$ is much more uniform.  
In the continuum Dirac field, the singular behavior around a 
more-than-six membered ring 
can be regarded as a consequence of the singular gauge fluxes, introduced in Ref. \onlinecite{Gonzalez93} 
for 2D cases, that effectively thread topological defects 
in the honeycomb lattice.  
Thus the wave functions and band structures presented here 
should also be reproduced in a continuum Dirac model 
with these gauge fluxes included.

Finally, it should be mentioned that
an electron confined to a curved graphene is generally affected by the local curvature
neglected in the present model,
where the hybridization between $sp_2$ and $sp_3$
occurs.\cite{Biase1994}
In real 3D graphite sponges, the curvature is expected to be concentrated 
around the topological defects, and it would give additional resonant states
as observed in the capped carbon nanotubes and the nanocones. \cite{Kim1999,Charlier2001}

{\it Conclusion---} 
We have studied the electronic structure of 3D P- and G-surface graphene networks with various length scales.
We find that the low-energy band structure 
is dominated by the topology of the structure, while 
within each topology 
the bands are basically determined 
by the periodicity $L$ in modulo 3.
The 3D graphene hosts massive Dirac fermions, 
with the mass scaled as $\sim 1/L$.
In the present study, we considered the model which only includes eight-membered rings as topological 
defects. Very recent works studied various 3D carbon networks that 
include seven-membered rings, and found metallic band structure.\cite{Weng2014,Tagami2014,commentmetallic} 
While their interest is focussed on 
smaller unit cell size $L\sim$ atomic scale than in the present paper, 
it would be interesting to study the behavior of those classes of 3D graphites 
to examine how the metallic bands would scale with larger $L$.

Let us make a few remarks on implications for experiments and future problems.
Unlike the present theoretical model, the experimentally realized 3D graphene network 
is at present not periodic \cite{Ito2014}, but has a randomly connected structure.  Still, we expect that 
singular wave amplitudes at the topological effects should be observed 
in the states around the charge neutrality point (i.e., when the 
system is not doped), as this is a topological property. 
The energy spectrum, on the other hand, should be different in the absence of 
periodicity, for which we would need some sort of the random network model
combined with the Dirac equation, especially to analyze 
whether a (smeared) gap can appear at the Dirac point as in the periodic case. 

The authors wish to thank Yoichi Tanabe, Yoshikazu Ito, and Mingwei Chen for helpful discussions and 
showing them some experimental results prior to publication.
This work was supported by JSPS Grants-in-Aid for Scientific Research (Grants Nos. 25107005, 26247064).

\end{document}